\documentstyle{article}

\font\tenrm=cmr10 \font\tenit=cmti10 \font\elevenbf=cmbx10
scaled\magstep 1 \font\elevenrm=cmr10 scaled\magstep 1
\newcommand{\be}{\begin{equation}}
\newcommand{\ee}{\end{equation}}

\newcommand{\bea}{\begin{eqnarray}}
\newcommand{\eea}{\end{eqnarray}}

\newcommand{\pr}{\partial}

\newcommand{\dg}{\dagger}

\input{epsf}
\renewenvironment{thebibliography}[1]
 { \elevenrm
   \begin{list}{\arabic{enumi}.}
    {\usecounter{enumi}     \setlength{\parsep}{0pt}
     \setlength{\itemsep}{3pt} \settowidth{\labelwidth}{#1.}
     \sloppy
    }}
     {\end{list}}
\begin{document}
\begin{center}
{\elevenbf Energy-Charge Dependence for Q-Balls}\\[0pt]
\vglue0.5cm {\tenrm T.A. Ioannidou\footnote{{\it Permanent
Address}: Institute of Mathematics, University of Kent, Canterbury
CT2 7NF, UK}$^{*}$, V.B. Kopeliovich$^{\dg}$, N.D.
Vlachos$^{\ddagger}$ } \\[0pt]
\vglue0.3cm $^{*}${\tenit Mathematics Division, School of
Technology,
University of Thessaloniki, Thessaloniki 54124, Greece}\\[0pt]
$^{\dg}${\tenit Institute for Nuclear Research of Russian Academy
of Sciences,
Moscow 117312, Russia}\\[0pt]
$^{\ddagger}${\tenit Physics Department, University of
Thessaloniki, Thessaloniki
54124, Greece }\\[0pt]
\vglue0.3cm {\it Emails: T.Ioannidou@ukc.ac.uk\\[0pt]
\hspace{12mm} vlachos@physics.auth.gr\\[0pt]
\hspace{16mm} kopelio@al20.inr.troitsk.ru}
\end{center}

\vglue0.3cm {\rightskip=2pc \leftskip=2pc \tenrm\baselineskip=11pt
\noindent We show that many numerically established properties of
Q-balls can be understood in terms of analytic approximations for
a certain type of potential. In particular, we derive an explicit
formula between the energy and the charge of the Q-ball valid for
a wide range of the charge }$Q$ .{\ \vglue0.6cm}
\elevenrm\baselineskip=14pt

\section{Introduction}

As shown by Coleman \cite{col}, the existence of Q-balls is a
general feature of scalar field theories carrying a conserved
$U(1)$ charge \cite{Lee}. Q-balls can be understood as bound
states of scalar particles and appear as stable classical
solutions (nontopological solitons) carrying a rotating time
dependent internal phase. They are characterized by a conserved
nontopological charge $Q$ (Noether charge) which is responsible
for their stability (see, for example, Refs.
\cite{drohm}-\cite{belova}). These features differentiate the
Q-ball interaction properties from those of the topological
solitons since here the charge $Q$ can take arbitrary values in a
specific range, allowing for the possibility of charge transfer
between solitons during the interaction process.

The concepts associated with Q-balls are extremely general and
occur in a wide variety of physical contexts \cite{hong}. Q-balls
are a generic consequence of the Minimal Supersymmetric Standard
Model (MSSM) \cite{mssm} where leptonic and baryonic balls may
exist. In this context the conserved charge is associated with the
$U(1)$ symmetries leading to baryon and lepton number
conservation, and the relevant $U(1)$ fields correspond to either
squark or slepton particles. Thus, the Q-balls can be thought of
as condensates of either squark or slepton particles. It has been
suggested that such condensates can affect baryogenesis via the
Affleck-Dine mechanism \cite{AD} during the post-inflationary
period of the early universe. Then, two interesting possibilities
occur: (i) If the Q-balls are stable and avoid evaporation into
lighter stable particles like protons, they are cosmologically
important since they can contribute to the dark matter content of
the universe \cite{dark}; (ii) If they are unstable, they decay in
a nontrivial way into baryons protecting them from erasure through
sphaleron transitions \cite{bar}.

Up till now, comprehensive studies of these objects have been made
by using either numerical simulations \cite{drohm}-\cite{AKP} or
some analytic considerations \cite{col,MV,PC}. In our approach we
will semi-analytically identify the explicit relation between the
energy and the charge of the Q-balls by using a semi-Bogomolny
argument in the energy density. Subsequently, we will show that
similar results can be obtained by using the Woods-Saxon ansatz
for describing the Q-ball profile function; an approach which has
been successfully applied to  describe analytically the properties
of multi-skyrmions in three \cite{K} and two spatial dimensions
\cite{IKZ}. This way, some universal properties of Q-balls in the
thin-wall limit can be established.

\section{Energy of the Q-balls}

Although Q-balls can exist in a variety of field theoretical models, we will
consider the $U(1)$ Goldstone model describing a single complex scalar field
$\phi $ in three spatial dimensions with potential $U(|\phi |)$. The
Lagrangian is
\begin{equation}
{\cal L}=\frac{1}{2}\partial _{\mu }\phi \,\partial ^{\mu
}\bar{\phi} -U(|\phi |)  \label{L}
\end{equation}
where the potential is only a function of $|\phi |$ and has a single minimum
at $\phi =0$. This is equivalent of stating that there is a sector scalar
particles (mesons) carrying $U(1)$ charge and having mass squared equal to
 $\frac{1}{2}U^{\prime \prime }(0)$. The corresponding energy functional is
given by
\begin{equation}
E=\int \left( \frac{1}{2}|\pr_t{\phi}|^{2}+\frac{1}{2}|\nabla \phi
|^{2}+U(|\phi |)\right) d^{3}x.  \label{En}
\end{equation}
The model has a global $U(1)$ symmetry leading to the conserved Noether
current
\begin{equation}
J_{\mu }=\frac{1}{2i}\left( \bar{\phi}\,\partial _{\mu }\phi -\phi
\,\partial _{\mu }\bar{\phi}\right).
\end{equation}
The conserved Noether charge $Q$ is
\begin{equation}
Q=\frac{1}{2i}\int \left( \bar{\phi}\,\partial _{t}\phi -\phi
\,\partial _{t}\bar{\phi}\right) d^{3}x.
\end{equation}
A stationary Q-ball solution has the form
\begin{equation}
\phi =e^{i\omega t}f(r)  \label{ph}
\end{equation}
where $f(r)$ is a real radial profile function which satisfies the ordinary
differential equation
\begin{equation}
\frac{d^{2}f}{dr^{2}}=-\frac{2}{r}\frac{df}{dr}-\omega
^{2}f+U^{\prime }(f) \label{gene}
\end{equation}
with boundary conditions $f(\infty )=0$ and $f^{\prime }(0)=0$.

This equation can either be interpreted as describing the motion of a point
particle moving in a potential with friction \cite{col}, or in terms of
Euclidean bounce solutions \cite{23}. In each case the effective potential
being $U_{eff}(f)=\omega^2f^2/2-U(f)$ leads to constraints on the potential
 $U(f)$ and the frequency $\omega$ in order for a Q-ball solution to exist.
Firstly, the effective mass of $f$ must be negative. If we
consider a potential $U(f)$ which is non-negative and satisfies
$U(0)=U^{\prime}(0)=0$, $U^{\prime\prime}(0)=\omega_{+}^2>0$ then
one can deduce that $\omega<\omega_{+}$. Furthermore, the minimum
of $U(f)/f^2$ must be attained at some positive value of $f$, say
$0<f_0<\infty$ and existence of the solution requires that
$\omega>\omega_{-}$ where $\omega_{-}^2=2U(f_0)/f_0^2$.
 Hence, Q-balls exist for all $\omega$ in the range $\omega_{-}<|\omega|<
\omega_{+}$.

Thus, the charge $Q$ of a stationary Q-ball solution simplifies to
\begin{eqnarray}
Q &\doteq &\omega I[f]  \nonumber \\
&=&4\pi \omega \int r^{2}f^{2}(r)\,dr  \label{Q}
\end{eqnarray}
where $I[f]$ is the moment of inertia. Numerical and analytical
methods have shown that when the internal frequency is close to
the minimal value $\omega _{-}$, the profile function is almost
constant, implying that the charge (\ref{Q}) is large (thin-wall
approximation). On the other hand, when the internal frequency
approaches the maximal value $\omega _{+}$ the profile function
falls off very quickly (thick-wall approximation). In the
thick-wall approximation the behavior of the charge $Q$ depends on
the particular form of the potential and the number of dimensions
\cite{PC}. In the case studied here we show that $Q\rightarrow
\infty $ as $\omega \rightarrow \omega _{+}$.

In order to derive the minimum of the energy of the Q-ball at
fixed charge $Q$, it is convenient to represent the energy in the
form
\begin{equation}
E_{_{{\tiny Q}}}=\frac{Q^{2}}{2I[f]}\,+4\pi \int \left(
\frac{f^{\prime }{}^{2}}{2}+U(f)\right) r^{2}\,dr  \label{ene}
\end{equation}
where the stabilizing role of the Q-ball rotational part is
obvious. Note that, without rotational energy the Q-ball would
collapse since $E_{_{Q}}\rightarrow 0$ as $f(r)\rightarrow 0$
everywhere except at the origin $r=0$. This is similar to the case
of rotating skyrmions when the 4-th order Skyrme term is omitted
in the Lagrangian. The zero mode (rotational) quantum correction
to the energy, which is proportional to $(I[f])^{-1}$ plays a
stabilizing role in this case.

The choice of the potential is not unique, the standard
requirement is that the function $U(f)/f^{2}$ has a local minimum
at some value of $f$ different from zero. Here we will consider
the following potential
\begin{equation}
U(f)=f^{2}\left[ 1+(1-f^{2})^{2}\right] .  \label{Uf}
\end{equation}
Note that $\omega _{+}=2$ and $\omega _{-}=\sqrt{2}$ and so that
stable Q-balls exist for $\sqrt{2}<\omega <2$.

\section{Semi-Bogomolny Argument}

We now proceed to obtain an ansatz for the profile function by
applying a semi-Bogomolny argument \cite{EB} in the energy
functional (\ref{ene}-\ref{Uf}). Initially, this approach was
applied to the Skyrme model \cite{Skyrme}, where it was shown that
the lower energy bound is proportional to the topological charge.
Although the model studied here is not a topological one (ie there
is no topological charge), the Bogomolny argument can still be
applied and leads to an upper energy bound. This way, the profile
function satisfies an exactly soluble first order differential
equation and the corresponding energy and charge density can be
easily derived. The same approach was applied for description of
multi-skyrmions properties, as presented in \cite{K,IKZ}.

The energy (\ref{ene}-\ref{Uf}) using the Bogomolny argument can be
expressed as:
\begin{eqnarray}
E_{_{B}} &=&\left( \frac{\omega }{2}+\frac{1}{\omega }\right) Q+4\pi \int
\left( \frac{1}{2}\left( f^{\prime}\right) ^{2}+f^{2}\left( 1-f^{2}\right)
^{2}\right) r^{2}dr  \nonumber \\
&=&\left( \frac{\omega }{2}+\frac{1}{\omega }\right) Q+4\pi \int \left(
\frac{f^{\prime }}{\sqrt{2}}+f\left( 1-f^{2}\right) \right)
^{2}r^{2}\,dr-4\pi \int \sqrt{2}f^{\prime }f\left( 1-f^{2}\right) r^{2}\,dr
\nonumber \\
&\geq &\left( \frac{\omega }{2}+\frac{1}{\omega }\right) Q-4\pi
\int \sqrt{2}f^{\prime }f\left( 1-f^{2}\right) r^{2}\,dr.
\label{EBog}
\end{eqnarray}

The equality is satisfied when the total square term is zero which gives the
semi-Bogomolny equation
\begin{equation}
f^{\prime }=-\sqrt{2}f(1-f^{2})\,.  \label{ode}
\end{equation}
The corresponding profile function has the simple form
\begin{equation}
f(r)=\frac{1}{\sqrt{1+C_{_{B}}\,\exp \left( 2\sqrt{2}\,r\right)
}}, \hspace{5mm}C_{_{B}}>0  \label{Bog}
\end{equation}
which satisfies the boundary conditions $f(0)=1/\sqrt{1+C_{_{B}}}$ and $%
f(\infty )=0$, while $f^{\prime
}(0)=-\sqrt{2}C_{_{B}}/(1+C_{_{B}})^{3/2}$. Note that the
asymptotic value of (\ref{Bog}): $f\sim \exp(-\sqrt{2}r)$ is in
agreement with the equation of motion  (\ref{gene}) for $\omega =
\sqrt{2}$ (ie for large values of $Q$). In fact, inside the
$Q$-ball $f' \sim 0$ and $f\sim 1$, while outside the $Q$-ball
$f'=\,0$ and $f=0$ and so (\ref{Bog}) describes accurately the
profile function of the Q-ball outside and inside its region where
the last term in (\ref{EBog}) vanishes. However, (\ref{Bog}) does
not describe accurately the profile function of the $Q$-ball on
its surface (the so-called transition region). Although, in the
thin-wall approximation the analytical and numerical results
(presented in Table 1) converge as $Q$ increases  since the
relative contribution of the surface decreases like $ Q^{-1/3}$ at
large $Q$.

For the specific form of the profile function (\ref{Bog}) the
charge and the energy can be evaluated explicitly. We find that
\begin{eqnarray}
Q_{_{B}} &\simeq &-\frac{\pi \omega }{2\sqrt{2}}\left( \frac{\pi
^{2}}{6} \,\ln \left( C_{_{B}}\right) +\frac{1}{6}\,[\ln \left(
C_{_{B}}\right)
]^{3}+Li_{3}\left[ -C_{_{B}}\right] \right) ,  \nonumber \\
E_{_{B}} &\simeq &\left( \frac{\omega }{2}+\frac{1}{\omega
}\right) Q+\frac{ \sqrt{2}\pi }{4}\left( \frac{\pi
^{2}}{6}+\frac{1}{2}\left[ \ln (C_{_{B}}) \right] ^{2}+\ln \left(
1+\frac{1}{C_{_{B}}}\right) +Li_{2}\left[ -C_{_{B}} \right]
\right)  \label{EQBog}
\end{eqnarray}
where $Li_{n}(z)=\int_{0}^{z}\frac{Li_{n-1}(y)}{{y}}\,dy$ and
$Li_{1}(y)=\ln (1-y)$ is the polylogarithm function.

Next, the equation for the energy $E_{_{B}}$ above has to be
minimized with respect to $C_{_{B}}$ while $Q_{_{B}}$ is kept
constant. We expect the semi-Bogomolny ansatz to be valid only
when $C_{_{B}}\approx 0$ where the initial ``velocity'' of the
trial profile function $f(r)$ tends to zero. In this region, the
logarithms will dominate the dilogarithm and the trilogarithm
functions respectively, since these functions tend to zero like
polynomials. Thus, by substituting $z=-\ln (C_{_{B}})$ for $z>0$
one obtains
\begin{eqnarray}
Q_{_{B}} &=&\frac{\pi \omega z}{12\sqrt{2}}\left( \pi ^{2}+z^{2}\right) ,
\nonumber \\
E_{_{B}} &=&\left( \frac{\omega }{2}+\frac{1}{\omega }\right)
Q_{_{B}}+\frac{ \sqrt{2}\pi }{4}\left( \frac{\pi
^{2}+3z^{2}}{6}+z\right) .  \label{EQ1}
\end{eqnarray}
One can explicitly solve $\omega $ in terms of the charge $Q_{_{B}}$ given
above in order to obtain that
\begin{equation}
\omega =\frac{12\sqrt{2}\,Q_{_{B}}}{\pi z\left( \pi
^{2}+z^{2}\right) }
\end{equation}
which when substituted into the energy gives
\begin{equation}
E_{_{B}}=\frac{6\sqrt{2}\,Q_{_{B}}^{2}}{\pi z\left( \pi
^{2}+z^{2}\right) }+ \frac{\pi z\left( \pi ^{2}+z^{2}\right)
}{12\sqrt{2}}+\frac{\sqrt{2}\pi }{4} \left( \frac{\pi
^{2}+3z^{2}}{6}+z\right) .
\end{equation}
Now, upon minimizing the energy with respect to $z$ we find that the
frequency $\omega $ and the charge $Q_{_{B}}$ are given by
\begin{eqnarray}
\omega ^{2} &=&2+\frac{12\,\left( 1+z\right) }{\pi ^{2}+3z^{2}},  \nonumber
\\
Q_{_{B}}^{2} &=&\frac{\pi ^{2}z^{2}\left( \pi ^{2}+z^{2}\right)
^{2}}{ 144\left( \pi ^{2}+3z^{2}\right) }\left( 6+6z+\pi
^{2}+3z^{2}\right) .
\end{eqnarray}
Finally, the relation between $z$ and $\omega $ is
\begin{equation}
z=\frac{2}{\left( \omega ^{2}-2\right) }\left( 1+\sqrt{1-\frac{\pi
^{2}}{12} \left( \omega ^{2}-2\right) ^{2}+\left( \omega
^{2}-2\right) }\,\right) . \label{z}
\end{equation}
It is obvious that in the limit $\omega \rightarrow \sqrt{2}$ the
parameter $z$ goes to infinity in consistency with the analytical
results (\ref{EQ1}). Thus, the semi-Bogomolny argument is valid in
the thin-wall approximation.

The elimination of the $z$ variable from the equations that define $Q_{_{B}}$
and $\omega $ can be easily performed. By letting $\varepsilon =\omega ^{2}-2
$ one gets
\begin{eqnarray}
z&=&\frac{2}{\varepsilon }\left( 1+\sqrt{1+\varepsilon -\frac{\pi ^{2}}{12}%
\,\varepsilon ^{2}}\,\right),\\
 Q_{_{B}} &=&\frac{\pi
\sqrt{2+\varepsilon }\left( 24+18\varepsilon +\left[
24+\varepsilon \left( 6+\pi ^{2}\varepsilon \right) \right] \sqrt{
1+\varepsilon -\frac{\pi ^{2}}{12}~\varepsilon ^{2}}\,\right)
}{9\sqrt{2}
\varepsilon ^{3}},  \label{Q1} \\[3mm]
E_{_{B}} &=&\frac{\pi \left( 2+\varepsilon \right) }{\sqrt{2}\,\varepsilon
^{2}}\sqrt{1+\varepsilon -\frac{\pi ^{2}}{12}~\varepsilon ^{2}}+\frac{\pi
\sqrt{2}\left( 1+\varepsilon \right) }{\varepsilon ^{2}}+\frac{4+\varepsilon
}{2\sqrt{2+\varepsilon }}\,Q_{_{B}}\ \cdot   \label{QAE}
\end{eqnarray}
It is obvious that $Q_{_{B}}\rightarrow \infty $ as $\omega
\rightarrow \sqrt{2}$ (which corresponds to $\varepsilon
\rightarrow 0$) while the upper energy limit for any value of $Q$
(or $\varepsilon $) can be obtained from (\ref{QAE}).

Next, by eliminating $\varepsilon $ between $Q_{_{B}}$ and
$E_{_{B}}$ we get the Q-ball energy-charge dependence since
\begin{eqnarray}
E_{_{B}}\!\!\!\! &=&\!\!\!\!\sqrt{2}\,Q_{_{B}}+\frac{3^{2/3}\pi
^{1/3}}{ 2^{7/6}}\,Q_{_{B}}^{2/3}+\frac{5\pi
^{2/3}}{2^{11/6}3^{2/3}}\,Q_{_{B}}^{1/3}- \frac{\pi \left( 4+3\pi
^{2}\right) }{36\sqrt{2}}+\frac{\pi ^{4/3}\left(
17-216\pi ^{2}\right) }{2592~2^{1/6}~3^{1/3}}\ Q_{_{B}}^{-1/3}  \nonumber \\
\!\!\!\! &&\!\!\!\!+\frac{\pi ^{5/3}\left( 20-54\pi ^{2}+27\pi
^{4}\right) }{ 1944~2^{5/6}~3^{2/3}}\,Q_{_{B}}^{-2/3}+O\left(
Q_{_{B}}^{-1}\right) . \label{EQ}
\end{eqnarray}
Note that for large values of $Q$ (ie for $Q^{1/3}\gg 1$) the
expression (\ref{EQ}) gives the upper energy bound. In Table 1 we
compare the results obtained from the semi-Bogomolny argument by
means of (\ref{EQ}) with the ones obtained by solving the full
second order differential equations (\ref{gene}) numerically, for
different values of $Q$.
 The agreement is impressive and appears to
extend far beyond the expected range of validity of the
semi-Bogomolny argument.

\begin{center}
\begin{tabular}{|l|l|l|l|l|l|} \hline
$Q$ & $E_{_{num}}$ & $E_{_{B}} $ & $\omega _{num}$ & $f(0)_{num}$
  \\ \hline 47.691 & 95.5007 & 87.9945 & 1.99750 & 0.1530  \\
\hline 25.703 & 51.6331 & 49.8406 & 1.98997 & 0.3020   \\ \hline
18.064 & 36.4665 & 36.0640 & 1.97737 & 0.4460   \\ \hline 15.168 &
30.7595 & 30.7182 & 1.95959 & 0.5798   \\ \hline 14.065 & 28.6076
& 28.6585 & 1.93649 & 0.7018   \\ \hline 14.193 & 28.8511 &
28.8982 & 1.90788 & 0.8093   \\ \hline 16.268 & 32.7688 & 32.7587
& 1.87350 & 0.9000   \\ \hline 19.256 & 38.2651 & 38.2417 &
1.83303 & 0.9723   \\ \hline 23.678 & 46.2186 & 46.2262 & 1.78606
& 1.0244   \\ \hline 34.910 & 65.9541 & 66.0215 & 1.73205 & 1.0555
  \\ \hline 61.603 & 111.220 & 111.482 & 1.67033 & 1.0654
  \\ \hline 149.263 & 253.840 & 254.515 & 1.60000 & 1.0564   \\
\hline 722.656 & 1140.119 & 1141.95 & 1.51987 & 1.0345 \\ \hline
\end{tabular}
\end{center}

Table 1: Comparison of the energy given by (\ref{EQ}) obtained
from the semi-Bogomolny argument ($E_{_{B}}$) with the ones
obtained by numerical simulations ($E_{num}$).\newline

 Note that although the semi-Bogomolny
argument and therefore the corresponding energy-charge formula
(\ref{EQ}) holds only for large values of $Q$ the agreement
between the numerical and analytical results holds for a wide
range of $Q$. However, although the profile (\ref{Bog}) describes
very well the energy of the Q-ball as a function of the charge
$Q$, its value at the origin $f(0)$ is always smaller than one and
differs considerably from $f(0)_{num}$ of Table 1.

\begin{figure}[tbp]
\begin{center}
\epsfxsize=9cm\epsfysize=9cm\epsffile{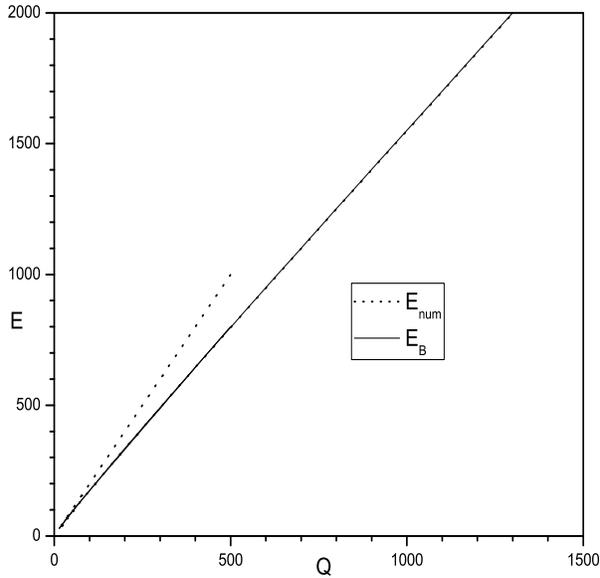}
\caption{Energy-charge dependence for $Q$-balls. The upper and
lower branches correspond to the thick-wall and thin-wall
approximations, respectively.} \label{fig-EQ}
\end{center}
\end{figure}

\begin{figure}[tbp]
\begin{center}
\epsfxsize=9cm\epsfysize=9cm\epsffile{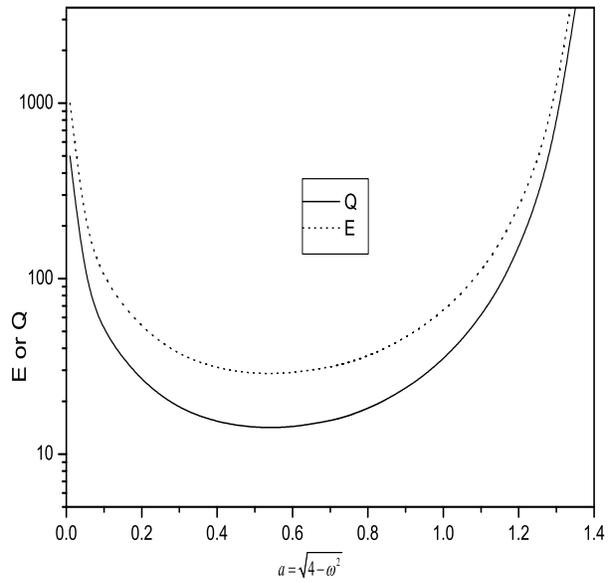} \caption{The energy
and charge (obtained numerically) as functions of the parameter
$\protect\alpha=\protect\sqrt{4-\protect\omega^2}$.}
\label{fig-EQ1}
\end{center}
\end{figure}

\begin{figure}[tbp]
\begin{center}
\epsfxsize=9cm\epsfysize=9cm\epsffile{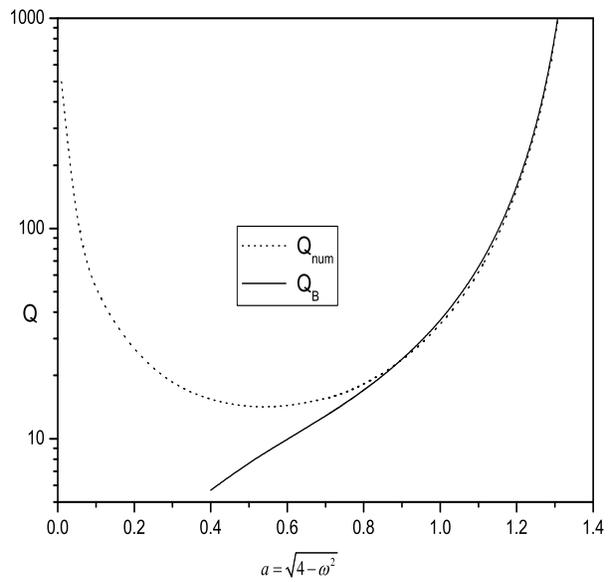 } \caption{The
charge $Q$ obtained numerically and
semi-analytically (\ref{Q1}%
) as a function of the parameter
$\protect\alpha=\protect\sqrt{4-\protect \omega^2}$.}
\label{fig-Q}
\end{center}
\end{figure}

In Figure \ref{fig-EQ} we plot the energy obtained from the
semi-Bogomolny argument (\ref{EQ}) and from the numerical
simulations ($E_{_{num}}$) for a wide range of $Q$. Note the
impressive agreement between the two results. The peculiar
behavior of Figure \ref{fig-EQ} for small $Q$ can be explained
from Figure \ref{fig-EQ1} where the energy and charge (obtained
numerically) as functions of $\omega$ are presented. It is obvious
that in a range of $\omega $ two different values of energy and
charge exist. Finally, in Figure \ref{fig-Q} we plot the charge
obtained from the semi-Bogomolny argument (\ref{Q1}) and from the
numerical simulations ($Q_{_{num}}$) in terms of the parameter
$a=\sqrt{4-\omega^2}$ in the allowed range $\sqrt{2}<\omega<2$.

{\it Remark:} The semi-Bogomolny  method can be applied for other
potentials of  polynomial type. For example, in \cite{MV} the
potential is of the form
 \be U(f)=f^2\left[1 + (1-f)^2\right]\label{Uff}
\ee and the corresponding semi-Bogomolny equation (analogous to
(\ref{ode})) is $f'=-f(1-f)$ while its solution
   \be
f(r)=\frac{1}{1+ \widetilde{C}_{_{B}}\,\exp(\sqrt{2}r)}.
\label{bb} \ee has the same asymptotic behavior for large $r$ as
in (\ref{Bog}).

\section{Woods-Saxon Ansatz}

A more general form of the ansatz (\ref{Bog}) is widely used in
nuclear physics describing nuclear matter distribution inside
heavy nuclei, or potential of nucleon-nucleus interaction. This is
the so-called Woods-Saxon distribution and the corresponding
profile function is given by
\begin{equation}
f(r)=\frac{C_{_{WS}}}{\sqrt{1+\exp \left[ a\left( r-r_{_{{\tiny
Q}}}\right) \right] }}.  \label{WS}
\end{equation}
In this case, three arbitrary parameters (instead of one) appear
in the parametrization, and thus, more degrees of freedom exist.
Here, $r_{_{Q}}$ corresponds to the radius of the Q-ball and $1/a$
defines the thickness of the shell of the Q-ball. At the origin,
the values of the profile function and its derivative are:
$f(0)=C_{_{WS}}\left( 1+e^{-ar_{_{Q}}}\right) ^{-1/2}$ and
$f^{\prime }(0)=-a\,C_{_{WS}}e^{-ar_{_{Q}}}\left(
2+2e^{-ar_{_{Q}}}\right) ^{-3/2}$. Recall that, due to boundary
conditions, $f^{\prime }(0)$ needs to be very small (in fact,
zero). This limit can be obtained when the product $ar_{_{Q}}$ is
large that is in the thin-wall approximation where $\omega
\rightarrow \sqrt{2}$ (see below). The field-theoretical
motivation for the Woods-Saxon ansatz was presented in the
previous section since for $C_{_{WS}}=1$ and $C_{_{B}}=\exp
(-ar_{_{Q}})$ the two profile functions given by (\ref{WS}) and
(\ref{Bog}) coincide.

For $Q$ large, the energy of the Q-balls given by
(\ref{ene}-\ref{Uf}) can be approximately evaluated using the
Woods-Saxon ansatz (\ref{WS}). To do so, integrals of the
following type are used:
\begin{eqnarray}
I_{n} &=&\int \frac{\left( C_{_{WS}}\right) ^{2n}}{\left( 1+\exp \left[
a(r-r_{_{{\tiny Q}}})\right] \right) ^{n}}\,r^{2}\,dr  \nonumber \\[3mm]
&=&\frac{1}{a^{3}}\,\left(C_{_{WS}}\right)^{2n}\left(
b^{2}I_{n}^{0}+2b\,I_{n}^{1}+I_{n}^{2}\right)  \label{In}
\end{eqnarray}
where $n=1,2,3$ and $b=ar_{_{Q}}$ while the moment of inertia
becomes: $ I[f]=4\pi I_{1}$. In general, by letting $w=\exp
[a(r-r_{_{Q}})]$ while $dr=dw/aw$,  the $m$-power of the integral
$I_n$ is defined as:
\begin{equation}
I_{n}^{m}=\int_{w_{0}}^{\infty }\frac{1}{z}\frac{(\ln
\,w)^{m}}{(1+w)^{n}} \,dw  \label{Imn}
\end{equation}
where $w_{0}=\exp (-b)$. After some algebra it can be easily shown that the
integrals $I_{n}^{m}$ are related via the following recursive relation
\begin{equation}
I_{n+1}^{m}=I_{n}^{m}-{\frac{m}{n}}I_{n}^{m-1}-\frac{(\ln
\,w_{0})^{m}}{ n(1+w_{0})^{n}}.
\end{equation}
For $w_{0}\ll 1$ one gets:
\begin{eqnarray}
&&I_{1}^{0}\simeq
b,\hspace{5mm}\hspace{5mm}\hspace{5mm}\hspace{5mm}\hspace{
5mm}\!\!I_{2}^{0}\simeq
b-1,\hspace{5mm}\hspace{5mm}\hspace{5mm}\hspace{5mm}
\!\!I_{3}^{0}\simeq b-{\frac{3}{2}},  \nonumber \\
&&I_{1}^{1}\simeq -{\frac{
b^{2}}{2}}+L,\hspace{5mm}\,\,\,\,\,\,I_{2}^{1}%
\simeq
-{\frac{b^{2}}{2}}+L,\hspace{5mm}\hspace{5mm}\,\,\,\,\,I_{3}^{1}
\simeq -{\frac{b^{2}}{2}}+L+{\frac{1}{2}},  \nonumber \\
&&I_{1}^{2}\simeq
{\frac{b^{3}}{3}},\hspace{5mm}\hspace{5mm}\hspace{5mm}
\hspace{5mm}\,\,I_{2}^{2}\simeq
{\frac{b^{3}}{3}}-2L,\hspace{5mm}\hspace{5mm}
\,\,\,\,\,I_{3}^{2}\simeq {\frac{b^{3}}{3}}-3L
\end{eqnarray}
where the parameter $L=2(1-1/4+1/9-1/16+...)\simeq 1.644$ does not
contribute in the leading order expansion of the parameter
$Q^{-1/3}$ (which we assume to be small -- see below). Then the
energy of the Q-ball (\ref{ene} -\ref{Uf}) can be approximated as
\begin{equation}
E_{_{Q}}\simeq \frac{Q^{2}}{\gamma 2VC_{_{WS}}^{2}}+\gamma
VC_{_{WS}}^{2} \bigl(2-\alpha 2C_{_{WS}}^{2}+\beta
C_{_{WS}}^{4}\bigr)+E_{deriv}
\end{equation}
where $\alpha =1-3b^{-1}$, $\beta =1-\frac{9}{2}b^{-1}$, $\gamma
=1+6Lb^{-2}$ and $V=\frac{4}{3}\pi r_{_{Q}}^{3}$. The derivative
term of the energy: $ E_{deriv}=\frac{1}{2}\int f^{\prime
}{}^{2}\,r^{2}\,dr$ is going to be neglected (initially).

The minimization of $E_{_{Q}}$ with respect to $\gamma
VC_{_{WS}}^2$ occurs at
\begin{equation}
(\gamma VC_{_{WS}}^2)_{_{min}}= \frac{Q}{\sqrt{2(2-2C_{_{WS}}^2
\alpha+C_{_{WS}}^4\beta)}}
\end{equation}
which determines $r_{_{Q}}$. The corresponding minimum of the energy is
\begin{equation}
E_{_{Q}}\simeq \sqrt{2}Q\left(2-2\alpha C_{_{WS}}^2 + \beta
C_{_{WS}}^4\right).
\end{equation}
Further minimization with respect to $C_{_{WS}}$ gives the following value
for the energy
\begin{equation}
E_{_{Q}}\simeq \sqrt{2}Q \biggl(2-{\frac{\alpha^2}{\beta}}\biggr)
\end{equation}
at $(C_{_{WS}})_{_{min}}^2=\alpha/ \beta$. For large $b$ (of order
$Q^{1/3}$), the parameter $(C_{_{WS}})_{_{min}}$ can be
approximated by $(C_{_{WS}})_{_{min}}\simeq
\sqrt{1+3/(2b)}=1+3/(4b)$ and in this limit the energy becomes:
\begin{equation}
E_{_{Q}}\simeq \sqrt{2}Q \left(1+{\frac{3}{4 b}}\right)  \label{EQB}
\end{equation}
where
\begin{equation}
(r_{_{Q}})_{_{min}} \simeq \biggl({\frac{3}{4\sqrt{2}\pi}}\biggr)
^{1/3}Q^{1/3}= 0.55 \,Q^{1/3}.  \label{rq}
\end{equation}
Note that equation (\ref{EQB}) implies that $E_{_{Q}} \rightarrow \sqrt{2}Q$
as $b \to \infty$.

Next the derivative energy contribution is considered:
\begin{equation}
E_{deriv}\simeq \frac{a\pi C_{_{WS}}}{4}\left(
r_{_{Q}}^{2}+\frac{2r_{_{Q}}}{a}+\frac{2L}{a^{2}}\right) .
\end{equation}
Taking into account the highest order terms in $r_{_{Q}}$ one obtains
\begin{equation}
E_{_{Q}}\simeq \sqrt{2}Q+{\frac{3Q}{2\sqrt{2}b}}+\frac{\pi
b(r_{_{Q}})_{_{min}}}{4}.
\end{equation}
Further minimization with respect to $b$ gives the energy-charge
dependence of the Q-balls up to order $O(Q^{2/3})$:
\begin{eqnarray}
E_{_{Q}} &\simeq &\sqrt{2}Q+\left( \frac{9\pi }{8\sqrt{2}}\right)
^{1/3}Q^{2/3}  \nonumber \\[3mm]
&=&\sqrt{2}\,Q+1.35702\,Q^{2/3}  \label{EQFIN}
\end{eqnarray}
which occurs when
\begin{eqnarray}
b_{_{min}} &=&\biggl({\frac{3\sqrt{2}Q}{\pi r_{_{{\tiny Q}}}}}\biggr)^{1/2}
\nonumber \\
&=&\left( \frac{12}{\pi }\,Q\right) ^{1/3}\simeq 1.563\,Q^{1/3}.
\label{bmin}
\end{eqnarray}
Note that the $O(Q^{2/3})$ terms of (\ref{EQFIN}) and (\ref{EQ})
coincide, and therefore the two analytic methods are in good
agreement. In addition, from (\ref{rq}) and since $b=ar_{_{Q}}$
one gets that $a_{_{min}}=2\sqrt{2}$ which is the value obtained
from the semi-Bogomolny argument (\ref{Bog}).

To conclude, let us state that in the thin-wall limit the profile
function (\ref{WS}) is given by $f(0)\sim 1+3/(4b)$ (or by
$f(0)\sim 1+1/(2b)$ including higher order corrections). Moreover,
in this limit the Woods-Saxon distribution (\ref{WS}) is close to
the exact values of the Q-ball profile function (obtained
numerically) presented in Table 1. In fact, for $Q=149.26$:
$f(0)=1.06$ and $f(0)_{_{num}}=1.0564$ while for $Q=722.66$:
$f(0)= 1.0356$ and $f(0)_{_{num}}=1.0345$.

{\it Remark:} For the alternative potential given by (\ref{Uff})
the semi-Bogomolny function (\ref{bb}) can be generalized in a
similar way as (\ref{WS}) and the corresponding energy is similar
to (\ref{EQFIN}) since
 \be E_{_{Q}}
\simeq \sqrt{2}Q+\left( \frac{\pi
}{3\sqrt{2}}\right)^{1/3}Q^{2/3}. \ee Note that the two
expressions differ only by a  $2/3$ factor before the term
$O(Q^{2/3})$.

\section{Discussion and Conclusions}

It has been shown that for specific parametrizations of the scalar
field a semi-analytic treatment for  Q-balls exists leading to
transparent and simple results. Two kinds of approximations have
been considered: one based on a semi-Bogomolny argument which
gives an exponential-step parametrization for the profile function
and the Woods-Saxon parametrization (the semi-Bogomolny
generalization) motivated also by nuclear physics experience.

The agreement of the results obtained using both approximations with the
numerical ones follows from the fact that the ansatz for the profile
function obtained from semi-analytic approaches have the correct asymptotic
behaviour as $Q$ and $r$ are large. This was not the case in the
semi-analytic treatment of the Skyrme model in three or two spatial
dimensions \cite{K,IKZ}. Although the energies obtained were accurate within
$0.5\%$ compared to the exact ones for large values of baryon number, the
asymptotic behaviour of the profile function was incorrect \cite{K}.

The thickness of the Q-ball surface (ie transition region) where
the profile $f$ decreases from $f \simeq 1$ to $f=0$ can be
estimated by
\begin{equation}
t\simeq {\frac{2}{b}}\,r_{_{Q}}.  \label{thick}
\end{equation}
For $Q$ large (and using the results of section 4), the thickness
is independent of the charge since $t\simeq \sqrt{2}$. Thus, the
large Q-balls can be visualized as spherically symmetric balls
with constant internal energy density
\begin{equation}
\rho _{_{E,V}}\simeq 2.
\end{equation}
These balls are surrounded by a surface of constant thickness and
constant average energy density per unit volume since
\begin{eqnarray}
\rho _{_{E,S}} &\simeq &\frac{E_{deriv}}{4\pi t\,r_{_{Q}}^{2}}
 \nonumber \\
&=&\frac{1}{4}
\end{eqnarray}
in natural units of the model. Therefore it is energetically
favorable for small Q-balls to fuse into a bigger one since the
surface of a single big Q-ball is smaller than the sum of surfaces
of several smaller Q-balls, for the same value of $Q$ (or with the
same total volume).

Our approach can be extended in lower (and also higher) spatial dimensions
in a natural way. In particular, in the one-dimensional case the
energy-charge dependence is
\begin{equation}
E(Q)_{_{1D}}\simeq \sqrt{2}Q+{\frac{1}{\sqrt{2}}}  \label{1}
\end{equation}
which is in a close agreement (within $1\%$) with the numerical
results obtained in \cite{BS}. In addition, the value of the
profile function at the origin is (approximately) given by
\begin{equation}
f(0)_{_{1D}}\simeq 1+{\frac{1}{4Q}}
\end{equation}
where terms of the form $\exp(-ar_{_{Q}})$ have been neglected
since $Q$ is large (ie $r_{_{Q}}>>$).

In the two-dimensional case the corresponding results are
\begin{equation}
E(Q)_{_{2D}}\simeq \sqrt{2}Q+\biggl({\frac{\pi
}{2\sqrt{2}}}\biggr)^{1/2} \sqrt{Q}  \label{2}
\end{equation}
while
\begin{equation}
f(0)_{_{2D}}\simeq 1+\left( \frac{\pi }{\sqrt{2}}\right)
^{1/2}\frac{1}{4 \sqrt{Q}}\,\cdot
\end{equation}

The formulas (\ref{1}), (\ref{2}) and (\ref{EQFIN}) indicate that
the relative contribution of the surface energy ($E_{deriv}$)
increases as the dimensionality of space increases. This property
of Q-balls can be useful in cosmological applications. In some
respect Q-balls are similar to the multiskyrmions which correspond
to bubbles of matter with universal properties of the shell, where
the mass and baryon number density is concentrated \cite{K,IKZ}.

We conclude by saying that our arguments can be extended to other
types of potentials.
Another important issue concerns the
semi-analytic identification of the profile function satisfying
equation (\ref{gene}) for (\ref{Uf}). We expect to report on this
problem soon.

\section*{Acknowledgements}

The work of VBK is supported by the Russian Foundation for Basic
Research, grant 01-02-16615. TI thanks the Royal Society and the
National Hellenic Research Foundation for a Study Visit grant.
\newline

{\elevenbf\noindent References} \vglue 0.1cm

\end{document}